\documentclass[letterpaper,twocolumn,british,english,prl]{revtex4-2}
\usepackage[utf8]{inputenc}
\usepackage[T1]{fontenc}
\usepackage{xcolor}
\usepackage{babel}
\usepackage{amsmath}
\usepackage{amssymb}
\usepackage{siunitx}
\usepackage{graphicx}
\usepackage[pdfusetitle,
 bookmarks=true,bookmarksnumbered=false,bookmarksopen=false,
 breaklinks=false,pdfborder={0 0 1},backref=false,colorlinks=true]
 {hyperref}
 \usepackage[dvipsnames]{xcolor}
\usepackage{hyperref}
 \hypersetup{
	pdftex,
	pdfstartview=FitH,
	bookmarksnumbered=true,
	pagebackref,
	colorlinks,
	breaklinks,
	urlcolor=Maroon,
	linkcolor=Maroon,
	citecolor=Bittersweet
}
\makeatletter

\pdfpageheight\paperheight
\pdfpagewidth\paperwidth

\usepackage{babel}
\usepackage{euler}

\makeatother

\begin{document}
\global\long\def\vect#1{\overrightarrow{\mathbf{#1}}}%

\global\long\def\abs#1{\left|#1\right|}%

\global\long\def\av#1{\left\langle #1\right\rangle }%

\global\long\def\imp#1{\left\langle #1\right\rangle _{\text{imp}}}%

\global\long\def\ket#1{\left|#1\right\rangle }%

\global\long\def\bra#1{\left\langle #1\right|}%

\global\long\def\tensorproduct{\otimes}%

\global\long\def\braket#1#2{\left\langle #1\mid#2\right\rangle }%

\global\long\def\bb#1{\boldsymbol{#1}}%

\global\long\def\b#1{\mathbf{#1}}%

\global\long\def\cal#1{\mathcal{#1}}%

\global\long\def\scr#1{\mathscr{#1}}%

\global\long\def\norm#1{\mathit{#1}}%

\global\long\def\braket#1#2{\left\langle #1|#2\right\rangle }%

\global\long\def\p#1{\left(#1\right)}%

\global\long\def\t#1{\text{#1}}%

\global\long\def\s#1{{\displaystyle {\displaystyle #1}}}%

\global\long\def\com#1#2{\left[#1,#2\right]}%

\global\long\def\an#1#2{\left\{  #1,#2\right\}  }%

\global\long\def\omv{\overrightarrow{\Omega}}%

\global\long\def\inf{\infty}%
\title{\selectlanguage{british}%
Markov Inequality as a Tool for Linear-Scaling Estimation \linebreak{}
 of Local Observables}
\author{\selectlanguage{british}%
H. P. Veiga}
\email{henriqueveigacj@gmail.com}
\address{\selectlanguage{british}%
Centro de Física das Universidades do Minho e Porto~~\\
 Departamento de Física e Astronomia, Faculdade de Ciências, Universidade
do Porto, 4169-007 Porto, Portugal}
\author{\selectlanguage{british}%
D. R. Pinheiro}
\address{\selectlanguage{british}%
Centro de Física das Universidades do Minho e Porto~~\\
 Departamento de Física e Astronomia, Faculdade de Ciências, Universidade
do Porto, 4169-007 Porto, Portugal}
\author{\selectlanguage{british}%
J. P. Santos Pires}
\address{\selectlanguage{british}%
Centro de Física das Universidades do Minho e Porto~~\\
 Departamento de Física e Astronomia, Faculdade de Ciências, Universidade
do Porto, 4169-007 Porto, Portugal}
\author{\selectlanguage{british}%
J. M. Viana Parente Lopes}
\email{jlopes@fc.up.pt}

\address{\selectlanguage{british}%
Centro de Física das Universidades do Minho e Porto~~\\
 Departamento de Física e Astronomia, Faculdade de Ciências, Universidade
do Porto, 4169-007 Porto, Portugal}
\date{\selectlanguage{british}%
\today}
\begin{abstract}
We introduce a linear-scaling stochastic method to compute real-space maps of any positive local spectral operator in a tight-binding model. By employing positive-definite estimators, the sampling error at each site can be rigorously bounded relative to the mean via the Markov inequality, overcoming the lack of self-averaging and enabling accurate estimates even under strong spatial fluctuations. The approach extends to non-diagonal observables, such as local currents, through local unitary transformations and its effectiveness is showcased by benchmark calculations in the disordered two-dimensional (2D) $\pi$-flux model, where the LDoS and steady-state current maps are computed. This method will enable simulations of disorder-driven mesoscopic phenomena in realistically large lattices and accelerate real-space self-consistent mean-field calculations.
\end{abstract}
\maketitle

Revealing spatially resolved quantum observables in large lattice systems remains a central challenge in condensed-matter theory. Quantities such as the LDoS, local currents, magnetization textures, and topological markers encode rich information about underlying quantum phenomena, including localization, Mott transitions, and quantum Hall effects. Understanding these observables is crucial not only for characterizing exotic phases of matter, but also for predicting how disorder, interactions, and topology interplay in realistic materials. Moreover, advancement of experimental techniques over the last 20 years -- e.g., \textit{scanning tunneling microscopy} (STM)\,\citep{brihuega_quasiparticle_2008,peres_local_2009,peres_scanning_2009}, scanning SQUID\,\citep{kirtley_fundamental_2010}, and \textit{nitrogen-vacancy center magnetometry}\,\citep{rondin_magnetometry_2014} -- have uncovered intricate spatial patterns in quantum materials, sparking the development of simulation methods capable of comparable resolution.

Spatial inhomogeneities are core to the physics of disordered systems. Anderson localization, for example, appears as an exponential suppression of wave-function amplitudes, which directly reflects in the LDoS\,\citep{evers_anderson_2008}, while their multifractal scaling near a metal-to-insulator (MI) transition is a clear signature of quantum criticality\,\citep{mirlin_distribution_1996,tran_statistics_2007,schubert_distribution_2010,mirlin_distribution_1994,malyshev_monitoring_2004,rodriguez_multifractal_2009,karcher_generalized_2022}. Likewise, observing local currents in a system can unveil robust edge channels for particle transport\,\citep{calogero_large-scale_2018,Nair2021}, signal presence or absence of backscattering\,\citep{konig_2013,rosiek_2023,Ivanov2024}, or even demonstrate the existence of vortices in steady-state transport regimes which had previously been connected to MI transition\,\citep{zhang_localization_2009,palm_observation_2024}. In addition, real-space topological markers, such as local Chern and $\mathbb{Z}_{2}$ invariants, have been used to detect topological order even in the absence of translational symmetry\,\citep{bianco_mapping_2011,varjas_computation_2020}.

Linear-scaling algorithms based on Chebyshev expansions are renowned as very efficient ways to estimate volume averaged spectral functions (i.e., operator traces) of sparse real-space Hamiltonians\,\citep{weisse_kernel_2006,ferreira_critical_2015,joao_kite_2020,joao_high-resolution_2022,GimenezDeCastro2024,Covaci2010}. While these can be adapted to compute local quantities, the generation of full-sample maps inevitably increases the computational effort to $\mathcal{O}(N^2)$ (N being the system size), as the use of stochastic evaluation becomes limited by the lack of self-averaging. This limitation is mainly significant for observables that fluctuate by orders of magnitude from point to point or if repeated sampling is required\,\citep{hutchinson_stochastic_1990, bekas_estimator_2007}.

In this paper, we introduce a new stochastic vector-based method that efficiently computes maps of positive local operators in large lattices, all at once and with controlled accuracy. Crucial to this approach is the use of positive-definite stochastic estimators, for which the celebrated \textit{Markov inequality} enables a rigorous bound of site-wise stochastic errors to a multiple of the local mean value, thus ensuring homogeneous convergence across the sample. Local unitary transformations naturally extend the approach to off-diagonal matrix elements.

The effectiveness of the method is showcased by benchmark calculations done on large samples of the disordered 2D $\pi$-flux model in the square lattice. First, vacancies are considered as the disorder source and full sample maps of the LDoS are computed. For a single vacancy at the center, the zero-energy LDoS is shown to decay as $r^{-2}$ away from the vacancy. This is an expected behavior caused by critical zero-energy states\,\citep{pereira_disorder_2006,ferreira_critical_2015}, but its numerical verification over several orders of magnitude is only accessible by the convergence properties of our method. For multiple vacancies, converged LDoS maps were obtained around zero-energy, demonstrating the method's ability to accurately capture details of the particle density field, even when it strongly fluctuates in space. Finally, steady-state current maps were computed in the presence of long-range potential scatterers. Their high spatial resolution reveals vortex-like patterns similar to those previously identified as hallmarks of the localization transition in graphene\,\citep{zhang_localization_2009}, belonging to the Kosterlitz–Thouless universality class.

\paragraph{Local Lattice Observables ---}\label{sec:Local-Observables}For independent quantum particles on a disordered lattice, much of the underlying physics is captured by real-space matrix elements of functions involving the single-particle Hamiltonian, $\mathcal H$. A central example is the spectral function,
\vspace{-0.05cm}
\begin{equation}
\mathcal{A}\left(\sigma',\mathbf{r}';\sigma,\mathbf{r}|\varepsilon\right)=\bra{\mathbf{r}',\sigma'}\delta\left(\varepsilon-\mathcal{H}\right)\ket{\mathbf{r},\sigma},\label{eq:LDoS-Definition}
\end{equation}
where $\varepsilon$ is the energy, $\left\{ \b r,\b r^{\prime}\right\}$ are lattice positions and $\left\{ \sigma,\sigma^{\prime}\right\}$ label orbitals within the unit-cell. This function reveals the spatial structure of the eigenstates of $\mathcal{H}$: it captures their static real-space properties around a given energy, and its diagonal elements yield the LDoS.
Off-diagonal elements, in turn, may also describe local dynamics—such as particle currents induced by an applied electric field. Specifically, the current from site $(\mathbf r,\sigma)$ to $(\mathbf r+\boldsymbol\Delta,\sigma')$ at time $t$ (in units of $2e/\hbar$) is
\vspace{-0.05cm}
\begin{equation}
I_{\b r,\b r+\b{\Delta}}^{\sigma\sigma^{\prime}}\left(t\right)
=\mathrm{Im}\!\left\{t_{\boldsymbol\Delta}^{\sigma\sigma'}\bra{\mathbf r,\sigma}\rho(t)\ket{\mathbf r+\boldsymbol\Delta,\sigma'}\right\},\label{eq:CurrentOverTime}
\end{equation}
where $\rho(t)=\cal{U}_t\rho_0\cal{U}_t^\dagger$ is the time-evolved single-particle density matrix and $t_{\boldsymbol\Delta}^{\sigma\sigma'}$ is the hopping parameter. For fermions, the equilibrium density matrix $\rho_{0}$ is written in terms of the Fermi-Dirac distribution, $f_{FD}\left(\mathcal{H}-\varepsilon_{F}\right)$\,\citep{veiga_unambiguous_2024}. 

As the method is not tailored to any specific observable, the presentation is kept general by considering real-space matrix elements of a Hermitian operator $X$. The diagonal element at site $\mathbf{r}$, denoted $x_{\mathbf{r}}=\bra{\mathbf{r}}X\ket{\mathbf{r}}$, is considered with the orbital index $\sigma$ omitted for brevity. In general, the full matrix representation of $X$ in the real-space basis is unavailable, so each $x_{\mathbf{r}}$ must be computed via $\mathcal{O}(N)$ matrix-vector multiplication, where $N$ is the Hilbert-space dimension.  
This way, obtaining the full spatial map of $x_{\mathbf{r}}$ would take $\mathcal{O}(N^{2})$ operations\,\cite{joao_kite_2020,joao_high-resolution_2022}. In contrast, the trace of large matrices (\textit{e.g.}, the sum over all $x_{\mathbf{r}}$) can be computed much more efficiently using stochastic sampling methods\,\citep{hutchinson_stochastic_1990, weisse_kernel_2006}. Stochastic trace estimation
starts by generating $R$ square-normalized vectors, $\left\{ \ket{\xi_{n}}\right\} _{n=1,\cdots,R}$, each
being a linear combination of basis states with
coefficients  $\xi_{n,\mathbf{r}}$. These coefficients are real or complex random variables such that $\overline{\xi_{n,\mathbf{r}}} = 0$ and $\overline{\xi^{*}_{n,\mathbf{r}}\xi_{n',\mathbf{r}'}} =\delta_{n,n'}\delta_{\mathbf{r},\mathbf{r}'}$ (complemented by $\overline{\xi_{n,\mathbf{r}}\xi_{n',\mathbf{r}'}} = 0$ in the complex case) where $\overline{\left(\dots\right)}$ is an average over independent random vectors. Following Ref.\,\citep{weisse_kernel_2006}, the trace of $X$ may be approximated as the average $\text{Tr}\left(X\right)\approx\frac{1}{R}\sum_{n=1}^{R}\bra{\xi_{n}}X\ket{\xi_{n}}$ which is expected to be self-averaging, that is, with a sampling error $\propto N^{-1}$.

Inspired by this algorithm, it is possible to recast the calculation of the full spatial map $x_{\mathbf{r}}$ as the average of a global stochastic estimator that samples all positions at once. A simple
unbiased estimator can be written as
\vspace{-0.1cm}
\begin{align}
\!\!\!\chi_{\mathbf{r}} & \!=\!\frac{1}{R}\sum_{n=1}^{R}\!\xi^{*}_{n,\mathbf{r}}\!\sum_{\mathbf{r}'}\!\bra{\mathbf{r}}\!X\!\ket{\mathbf{r}'}\xi_{n,\mathbf{r}'}\!=\frac{1}{R}\sum_{n}\!\bra{\xi_{n}}\!\circ\!X\!\ket{\xi_{n}} \label{naive_estimator},
\end{align}
with the notation $\bra{\psi_{1}}\circ\ket{\psi_{2}}=\psi_{1}^{*}\left(\mathbf{r}\right)\psi_{2}\left(\mathbf{r}\right)$, representing the (element-wise) Hadamard product. Although Eq.\,\eqref{naive_estimator} represents a legitimate estimator ($\overline{\chi_{\mathbf r}}=x_{\mathbf r}$), a subtle but important issue remains regarding its convergence properties. The sample-to-sample fluctuations of this estimator are quantified by its variance,
\vspace{-0.05cm}
\begin{equation}\label{eq:Variance_NaiveAlgorithm}
\sigma_{\mathbf r}^{2} = [X^{2}]_{\mathbf r,\mathbf r} + \Xi[\xi]\,X_{\mathbf r,\mathbf r}^{2},
\qquad
\Xi[\xi]=\overline{|\xi|^{4}}-2, 
\end{equation}
where $\overline{|\xi|^{4}}$ is the $4^{\text{th}}$-moment of the random-vector component distribution. Specifically, in the eigenbasis of $X$, the relative sampling error takes the form of
\vspace{-0.05cm}
\begin{equation} 
\frac{\sigma_{\b r}}{X_{\b r,\b r}}=\frac{\sqrt{\sum_{\alpha}\left[X^{2}\right]{}_{\alpha,\alpha}\left|\psi_{\alpha}\left(\b r\right)\right|^{2}}}{\sum_{\alpha}X{}_{\alpha,\alpha}\left|\psi_{\alpha}\left(\b r\right)\right|^{2}}, \label{eq:Local-Variance-Eigenstate}
\end{equation}
revealing its dependence on the spatial structure of the wavefunctions. If $X$ (as in Eq.\,\eqref{eq:LDoS-Definition}) selects a narrow spectral window that effectively isolates a single eigenstate $\psi_{\alpha}(\mathbf r)$, the relative local error scales as $|\psi_{\alpha}(\mathbf r)|^{-1}$. This results in a non-uniform spatial convergence, preventing an accurate LDoS estimation with feasible computational time in localized systems where wavefunction amplitudes can vary by orders of magnitude in space.

\paragraph{Positive-Definite Estimator ---}The spatially uneven convergence can be eliminated by using an estimator whose sampling noise scales with the local mean. For positive-definite operators ($X_{\mathbf r}\!\ge0$), this can be achieved by forcing the samples to be non-negative, in contrast to the sign-changing ones produced by Eq.\,\eqref{naive_estimator}. If a random field $\eta_{\mathbf{r}}$ is non-negative, the \textit{Markov inequality} bounds
\vspace{-0.05cm}
\begin{equation}
P(\eta_{\mathbf{r}}\geq C)\leq\overline{\eta_{\mathbf{r}}}C^{-1}
\end{equation}
for any $C>0$. Consequently, the $\alpha^{\text{th}}$ quantile is at most a multiple of its mean ($q_\alpha\le (1-\alpha)^{-1}\overline{\eta_{\mathbf r}}$) implying that the relative fluctuations of remain constant in space.

To exploit this property, a positive-definite unbiased estimator is constructed for the diagonal elements of the operator \(X_{\mathbf r}\). A new random field \(\phi_{\mathbf r} = \langle \xi | \circ \sqrt{X} | \xi \rangle\) is introduced so that the target quantity \(X_{\mathbf r}\) can be obtained either as \(\overline{|\phi_{\mathbf r}|^{2}}\) for complex random vectors or as \(\mathrm{Var}(\phi_{\mathbf r})\) for real random vectors satisfying \(\overline{|\xi|^{4}}=2\), as follows directly from Eq.~\eqref{eq:Variance_NaiveAlgorithm}. In both cases, the field $X_{\b r}$ is estimated by non-negative sampling which leverages Markov Inequality and ensures controlled relative errors. The estimators are also basis-independent: if the basis \(\{\ket{\mathbf r}\}\) is replaced by a unitarily transformed basis \(\{\ket{\mathbf u}\} = \{U\ket{\mathbf r}\}\), the corresponding random field $\left\{\phi_{\b u}\right\}$ can be likewise used to estimate diagonal elements in that basis. The reasoning also applies to the unitary time-evolution of Eq.\eqref{eq:CurrentOverTime}.

\begin{figure}[t]
\centering{}\includegraphics[scale=0.55]{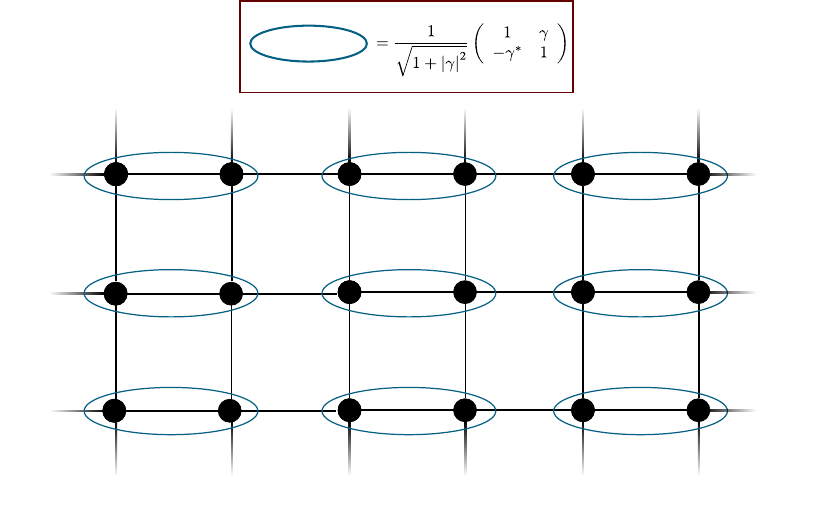} \vspace{-0.4cm}\caption{\foreignlanguage{english}{
Schematic representation of a bipartite squared lattice. The blue
ellipses represent connected pairs. These form a complete set, and
for each pair we apply the change of basis matrix shown in the red
box. \vspace{-0.3cm}}}\label{fig:Lattice Bipartition}
\end{figure}

\paragraph{Off-Diagonal Generalization ---}\label{sec:Estimating-Off-Diagonal-Elements}
Even though the method had initially been formulated for the estimation of diagonal matrix elements, it can be naturally extended to evaluate off-diagonal elements (as in Eq.\,\eqref{eq:CurrentOverTime}). The key idea is that diagonal elements in a suitably rotated basis encode information about the corresponding off-diagonal terms. Specifically, if one applies a local unitary transformation to a pair of sites $\left\{\b r,\b r^{\prime}\right\}$ of the form:
\vspace{-0.1cm}
\begin{equation}
\begin{cases}
\ket{\alpha_{\gamma}}=\frac{1}{\sqrt{1+\abs{\gamma}^{2}}}\left(\ket{\mathbf{r}}+\gamma\ket{\mathbf{r}'}\right)\\
\ket{\beta_{\gamma}}=\frac{1}{\sqrt{1+\abs{\gamma}^{2}}}\left(-\gamma^{*}\ket{\mathbf{r}}+\ket{\mathbf{r}'}\right),
\end{cases}
\end{equation}
where $\gamma\in\mathbb{C}$ is a parameter controlling the rotation, the diagonal elements in the rotated basis will be related to the original off-diagonal matrix elements in as follows,
\begin{align}
\s{2\bra{\b r}X\ket{\b r^{\prime}}\,=\,\Upsilon_{1}-i\Upsilon_{i}},
\end{align}
where $\Upsilon_{\gamma}=\bra{\alpha_{\gamma}}X\ket{\alpha_{\gamma}}-\bra{\beta_{\gamma}}X\ket{\beta_{\gamma}}$. Importantly, because
each bond rotation acts only on disjoint site pairs and the different components of the random vector are uncorrelated,
the procedure can be applied to all such bonds in parallel
without inducing additional sampling correlations (illustrated in
Fig.$\,$\ref{fig:Lattice Bipartition}). Hence, by iterating through the different groupings needed to cover all inter-sub-lattice
bonds, this method yields a complete map of the off-diagonal structure
of $X$.

\begin{figure}[t]
  \centering
  \hspace{-1.0cm}\includegraphics[scale=0.25]{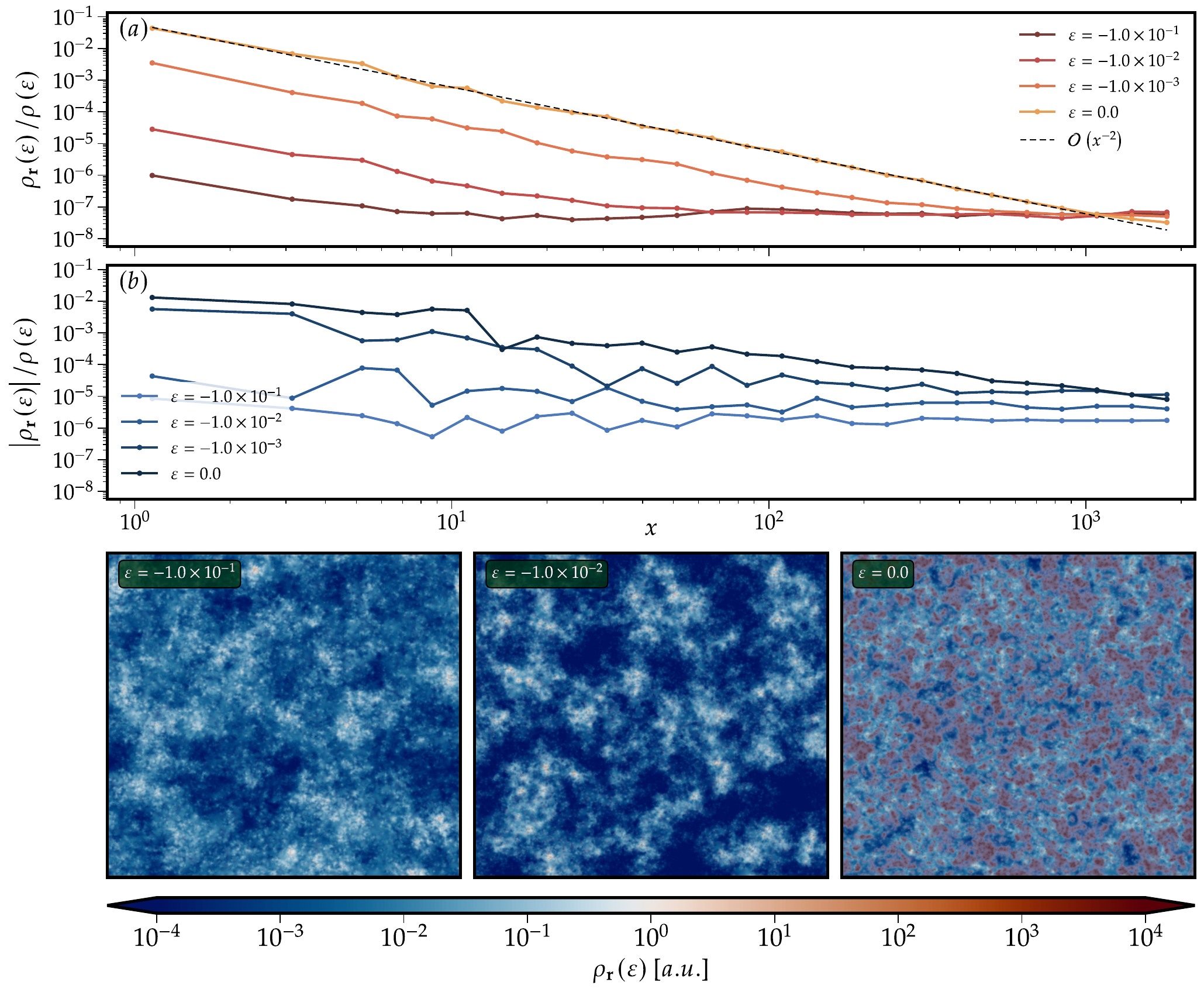}
  \vspace{-0.2cm}
  \caption{(a) Longitudinal cut across a single central vacancy on a $4096\times4096$ supercell computed with the positive-definite estimator ($32$ random vectors), showing the expected power-law decay as $\varepsilon\to0$. (b) Same cut but computed with the conventional estimator, which yields both non-positive values and misses small amplitudes. Bottom panels: LDoS maps for a $2048\times2048$ supercell with 0.5\% vacancies at three representative energies and fixed spectral resolution of $100 \si{\micro\electronvolt}$.}
  \label{fig:results-LDoS}
  \vspace{-0.3cm}
\end{figure}

\paragraph{Demonstration 1: Electron Density Fields ---}\label{sec:ApplicationPiFlux}
The 2D $\pi$-flux model on a square lattice is used as a test bed for the method. Its initially pristine periodic Hamiltonian
\begin{equation}
\cal H_{\b k}\,=\,-2w\left(\cos k_{x}\sigma_{x}+\cos k_{y}\sigma_{z}\right),
\end{equation}
where $w$ is the hopping parameter, loses its translation symmetry due to the presence of point defects (vacancies). As a first demonstration, we test our method by unveiling this zero-energy mode for a single central vacancy in a $4096\times4096$ supercell under twisted boundary conditions. The LDoS, $\rho_{\mathbf{r}}(\varepsilon)$, was computed along a longitudinal cut starting at the vacancy for a set of energies approaching $\varepsilon=0$, using KPM with a spectral resolution of $1 \si{\milli\electronvolt} $. Random sampling was performed with both the proposed positive-definite estimator and the conventional mean estimator, using the same ensemble of $32$ random vectors to compare convergence behaviour. Figures \ref{fig:results-LDoS} (a) and (b) show that the positive-definite estimator faithfully captures the algebraic decay of the zero-energy LDoS, $\rho_{\mathbf{r}}(0)\propto|\mathbf{r}|^{-2}$, over several orders of magnitude, whereas the naive estimator does not. The latter fails in regions of very low LDoS, where sampling noise generates nonphysical negative values and large relative errors—an instance of a random-sampling \textit{sign problem} that the positive-definite formulation inherently avoids.

For a finite vacancy concentration (0.5\% on a $2048\times2048$ supercell)
the real-space LDoS maps are feature-rich. Computing them with a fixed
$100 \si{\micro\electronvolt}$ spectral resolution, averaging over $1000$ random vector
and twist-angle configurations, we demonstrate the controlled accuracy
of the positive-definite estimator. Despite the LDoS maps spanning
several orders of magnitude the 95th percentile of the local
relative fluctuations is upper bounded at 3.5\%. The final panel
highlights the overlap of the zero energy critical states, with
local amplitudes at least two orders of magnitude larger than the
noncritical regions.

Next, we analyze a $512\times512$ sample with 1\% randomly placed vacancies at $\varepsilon = 0$ with
a $200 \si{\micro\electronvolt} $ spectral resolution. To quantify the
accuracy of both estimators we measure the local relative error
\vspace{-0.05cm}
\begin{equation}
\s{\Delta_{\b r}\,=\,\frac{\left|\rho_{\b r}^{\t{St}}\left(\varepsilon\right)-\rho_{\b r}^{\t{Ex}}\left(\varepsilon\right)\right|}{\rho_{\b r}^{\t{Ex}}\left(\varepsilon\right)},}
\end{equation}
where $\rho_{\b r}^{\t{Ex}}$ is the exact LDoS and $\rho_{\b r}^{\t{St}}$ is the estimate.

\begin{figure}[t]
  \centering
  \hspace{-0.4cm}\includegraphics[scale=0.28]{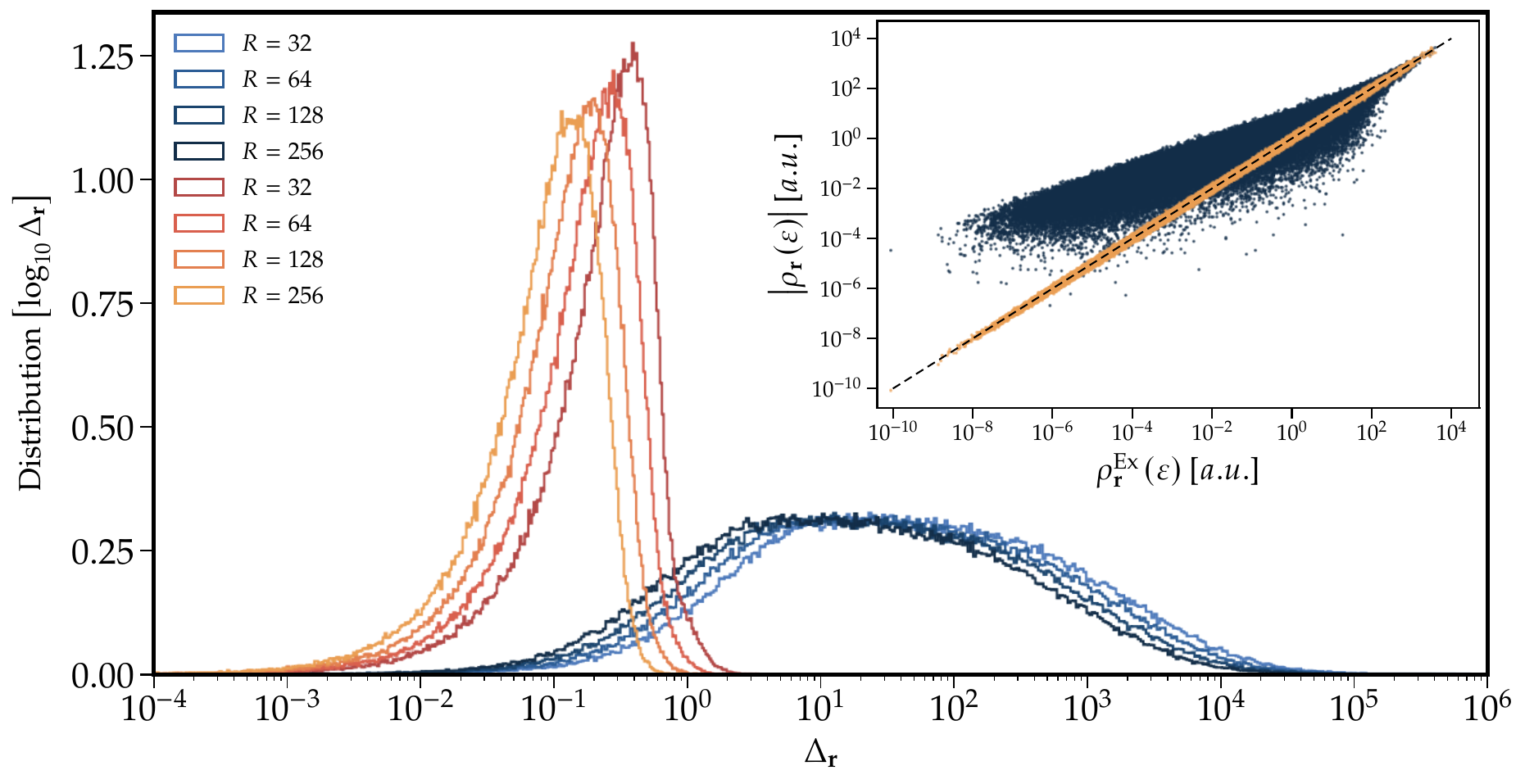} 
  \vspace{-0.2cm}
  \caption{Probability density functions of the local relative error $\Delta_{\b r}$ for the
  conventional (shades of blue) and positive-definite (shades of red) stochastic
  estimators, as a function of the number of random vectors, $R$. The latter
  reduces the mode of $\Delta_{\b r}$ by approximately two orders of magnitude, while
  scaling as $R^{-0.5}$. The inset is a scatter plot of stochastic and exact LDoS for $256$ random vectors.
  \vspace{-0.3cm}}
  \label{fig:Error-LDoS}
\end{figure}

Figure $\!$\ref{fig:Error-LDoS} demonstrates that the positive-defined estimator (shades of red)
dramatically reduces relative errors. The corresponding statistical distribution mode
lies about two orders of magnitude below that of the naive estimator (shades of blue), which
also exhibits a much broader spread. The first moment of the error distribution follows the expected $R^{-1/2}$ scaling with the number of random vectors $R$. The inset shows the
local observable scatterplot comparing
stochastic estimates with the exact LDoS maps ($256$ random vectors). The
naive estimator (blue) fails to reproduce values below approximately $10^{2}$. Furthermore,
it produces nonphysical negative LDoS (here plotted as absolute values). Contrastingly,
the positive-definite estimator (yellow) exhibits a controlled dispersion for over ten orders of magnitude. Moreover, for the highest $R$, each stochastic estimation took $\approx$ 4 \text{cpu.h}, whereas the exact result took $\approx$ 171 cpu.days.

\begin{figure}[t]
  \centering
  \vspace{-0.9cm}
  \includegraphics[scale=0.345]{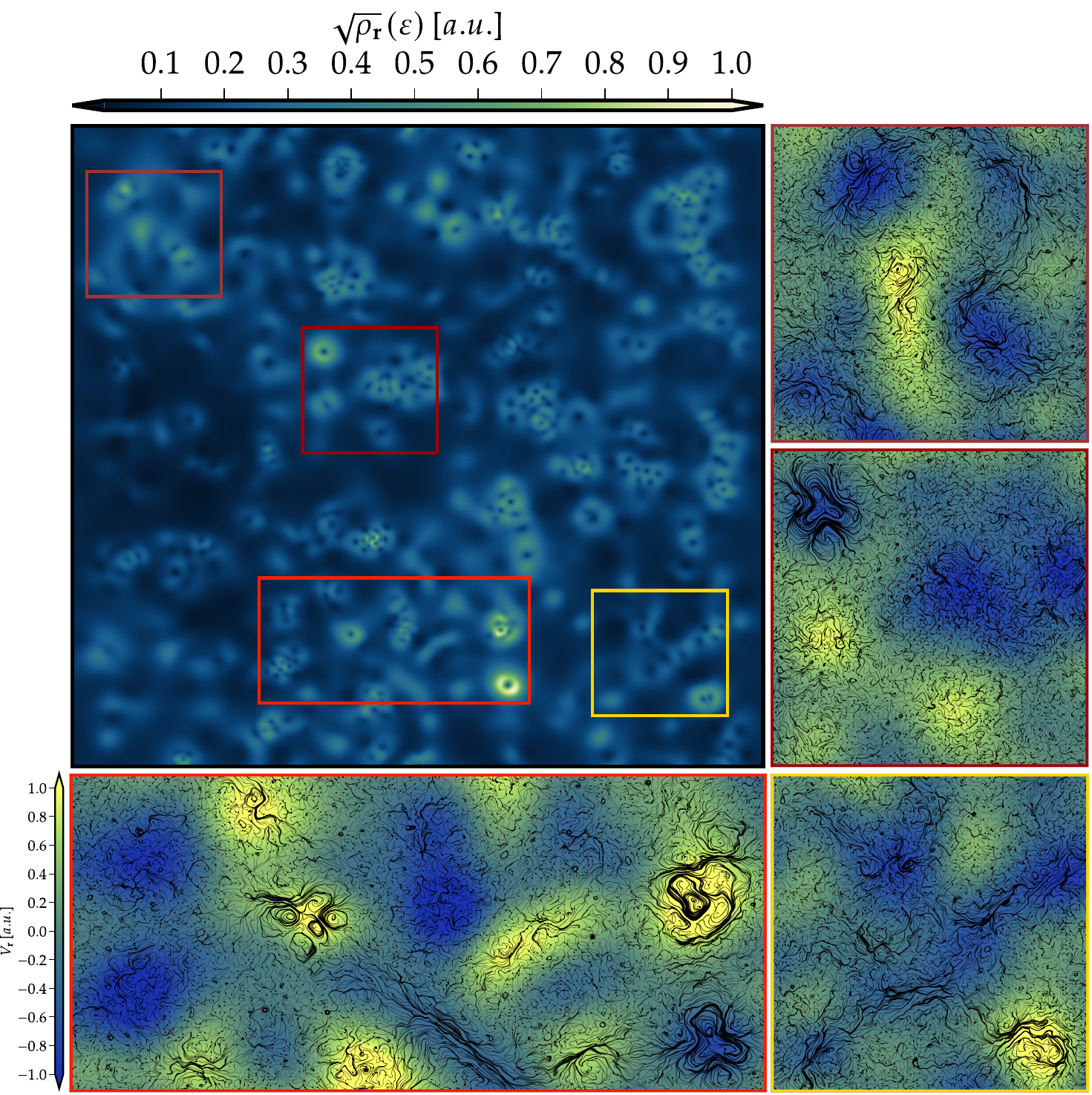} 
 \caption{Square-root LDoS and steady-state local currents for a $512\times512$ sample in the two-terminal setup with 0.5\% concentration of long-range impurities ($W=1.1 \si{\electronvolt}$, $\kappa=9.0$). Top: square-root LDoS map ($500$ random vectors and spectral resolution $100 \si{\micro\electronvolt}$). Magnified regions demonstrate the local currents steady state fields. Intricate patterns in the real-space distributions of currents is evident with vortex–anti-vortex pairs being visible.
 \vspace{-0.3cm}
 }
  \label{fig:results-Currents}
\end{figure}

\paragraph{Demonstration 2: Local Currents Fields ---} To illustrate the versatility
of this method we studied the local current fields that develop when
translation symmetry is broken by strong long-range impurities. The disordered potential
is given by $V_{\b r}\,=\,\sum_{m=1}^{M}w_{m}\exp\left(-0.5\left|\b r-\b r_{m}\right|^{2}/\kappa^{2}\right)$, where the scatterer amplitudes $w_{m}$
are drawn uniformly from $\frac{W}{2}\com{-1}1$. We analyse a $512\times 512$
sample with $0.5\%$ randomly placed impurities with
$W=1.1 \si{\electronvolt}$ and $\kappa=9.0$. This sample
is connected to two finite pristine leads (each $512 \times 512$) and periodic boundary
conditions are imposed in the transverse direction.
Following the linear-scaling
time-evolution methodology described in
\citep{santos_pires_landauer_2020,veiga_unambiguous_2024} we compute the real space distribution
of the steady-state local current fields. The LDoS shown in
Figure$\,$\ref{fig:results-Currents} was obtained using $500$ random vectors. The spectral
probe is centered on the Fermi energy $\varepsilon_{F}=0.1\si{\electronvolt}$ and has a width of $100 \si{\micro\electronvolt}$. In terms of vacancy calculations, the LDoS exhibits controlled
accuracy. The 95th percentile of local relative fluctuations is $\lesssim 5\%$. The statistical properties of the diagonal matrix elements carry over to the matrix elements in the rotated basis. However, the relation between the magnitudes of the individual matrix elements and their differences
is model dependent, so it is challenging to draw general conclusions for off-diagonal estimates.
The simulation of the linear response steady state current fields is performed by exploiting statistical correlations. Both the equilibrium and steady-state fields are computed with
the same random vector. Subtraction of both estimates with random vectors $2.4\times10^{4}$ reveals intricate structures,
with tightly bound vortex–anti-vortex pairs and complex flow patterns that appear only when local currents are resolved at this scale.

\paragraph{Conclusions ---}\label{sec:Conclusions}To summarize, we introduced a new linear-scaling stochastic method to compute full-sample maps of positive local spectral operators in tight-binding systems. By employing positive-definite estimators, the sampling error at each site is bounded relative to its mean value via the Markov inequality, ensuring homogeneous sampling convergence across the sample even when the observable fluctuates strongly from point to point. This precise control of the site-wise error is crucial, since unlike stochastic trace evaluations, random sampling of local quantities lacks self-averaging. The method was further extended to evaluate non-diagonal observables (e.g, local currents) by means of local unitary transformations.

The method was illustrated using three representative cases in the 2D $\pi$-flux model. For a single vacancy, we computed the real-space LDoS and recovered the expected zero-energy $r^{-2}$ decay\,\citep{pereira_disorder_2006,ferreira_critical_2015}, demonstrating that our approach can accurately capture the converged tail over several orders of magnitude — something not attainable with a naive mean estimator. Moreover, full-sample LDoS maps near the nodal energy were obtained, maintaining a controlled accuracy even when vacancy-induced critical states dominate and drive fast spatial variations spanning several orders of magnitude. Finally, for the $\pi$-flux model with Gaussian potential scatterers, we computed steady-state current maps in a two-terminal setup using the time-evolution approach introduced in Refs.\,\citep{santos_pires_landauer_2020,veiga_unambiguous_2024}. The resulting current patterns reveal subtle vortex structures akin to those previously associated with the metal-to-insulator transition in graphene\,\citep{zhang_localization_2009}, belonging to the Kosterlitz–Thouless universality class.

\paragraph{Outlook ---}\label{sec:Outlook}Our proposal opens up new possibilities for large-scale simulations of mesoscopic phenomena aided by a reduction in computational complexity, from \textit{quadratic} to \textit{linear} in the lattice size. This enables the measurement of local quantities across entire samples, which can be applied, for example, to characterize multifractality of disordered phases\,\citep{mirlin_distribution_1996,tran_statistics_2007,schubert_distribution_2010,goncalves_disorder-driven_2020,stosiek_multifractal_2021} or to study disorder-driven transitions\,\citep{mirlin_distribution_1994,malyshev_monitoring_2004,evers_anderson_2008,rodriguez_multifractal_2009,pixley_rare-region-induced_2016,buchhold_vanishing_2018,pires_breakdown_2021,santos_pires_anomalous_2022,karcher_generalized_2022}. In addition, its efficiency and precision make it suitable for use in self-consistent mean-field approaches to interactions in real-space\,\citep{joao_kite_2020, joao_high-resolution_2022}, and the positive-estimator principle can further improve evaluation of other global quantities.

This work was supported by Fundação para a Ciência e a Tecnologia
(FCT, Portugal) in the framework of the Strategic Funding UIDB/04650 - Centro de Física das Universidades do Minho e do Porto. Further support from FCT through Projects No. POCI-01-0145-FEDER028887 (J.M.V.P.L.) and
PhD Grant. No. 2024.00560.BD (H.P.V.) are acknowledged. The authors
acknowledge fruitful discussions with Aires Ferreira, Bruno Amorim,
Caio H. Lewenkopf, Vitor M. Pereira, J. M. Alendouro Pinho, J. M.
B. Lopes dos Santos, and Simão M. João.
\vspace{-0.4cm}
\bibliographystyle{apsrev4-2}
\bibliography{References}
\selectlanguage{english}%

\end{document}